\documentclass[preprint,12pt]{elsarticle}

\usepackage{graphicx}
\usepackage{float}
\usepackage{amsmath}
\usepackage{amssymb}
\usepackage{color}

\journal{J. Electron Spectroscopy Rel. Phenom.}

\begin{document}

\begin{frontmatter}

\title{Magnetic dichroism in angular resolved hard X-ray photoelectron spectroscopy from
       buried magnetic layers.}

\author{Carlos E. ViolBarbosa}

\author{Siham Ouardi}

\author{Gerhard H. Fecher \corref{cor}}
\cortext[cor]{Corresponding author: Gerhard H. Fecher, fecher@cpfs.mpg.de}

\author{Daniel Ebke}

\author{Claudia Felser}

\address{Max Planck Institute for Chemical Physics of Solids,
             N{\"o}thnitzer Str. 40, 01187 Dresden, Germany}

\begin{abstract}

This work reports on the measurement of magnetic dichroism in angular-resolved
photoelectron spectroscopy from in-plane magnetized buried thin films. The high
bulk sensitivity of hard X-ray photoelectron spectroscopy (HAXPES) in
combination with circularly polarized radiation enables the investigation of the
magnetic properties of buried layers. Angular distributions of high kinetic energy (7 to
8~keV) photoelectrons in a range of about $60^\circ$ were recorded in parallel
to the energy distribution. Depending on purpose, energy and angular resolutions
of 150 to 250~meV and $0.17^\circ$ to $2^\circ$ can be accomplished
simultaneously in such experiments. Experiments were performed on
exchange-biased magnetic layers covered by thin oxide films. More specifically, the
angular distribution of photoelectrons from the ferromagnetic layer Co$_2$FeAl
layer grown on MnIr exchange-biasing layer was investigated where the magnetic
structure is buried beneath a MgO layer. Pronounced magnetic dichroism is found
in the Co and Fe $2p$ states for all angles of emission. A slightly increased
magnetic dichroism was observed for normal emission in agreement with
theoretical considerations.

\end{abstract}


\end{frontmatter}

\section{Introduction} 

Magnetic circular dichroism (MCD) in photoabsorption and photoemission has
become a very powerful tool for the element-specific investigation of the
magnetic properties of alloys and compounds. Thus far, such studies have been
mainly carried out using soft X-rays, resulting in a rather surface sensitive
technique due to the low electron mean free path of the resulting low energy
electrons. The application of hard X-rays~\cite{Lind74} results in the emission
of electrons with high kinetic energies allowing an increase of probing
depth~\cite{Koba03}. For $h\nu>8$~keV, the bulk spectral weight was found to
reach more than 95\%~\cite{Suga09}. Hard X-ray photoelectron spectroscopy
(HAXPES) has been found to be a well-adaptable non-destructive technique for the
analysis of chemical and electronic states~\cite{FBG08,Koz10}. It was recently
shown that HAXPES can be combined easily with variable photon polarization when
using phase retarders~\cite{Ueda08}. Linear dichroism in the angular
distribution of the photoelectrons is achieved using linearly polarized hard
X-rays and is succesfully applied to identify the symmetry of valence band states
in Heusler compounds~\cite{Ouardi11}. In combination with excitation by
circularly polarized X-rays~\cite{Ueda08}, this method will serve as a unique
tool for the investigation of  magnetoelectronic properties of deeply buried
layers and interfaces in magnetic multilayer structures.

Baumgarten {\it et al}~\cite{Baum90} carried out a pioneering study on magnetic
dichroism in photoemission and observed this phenomenon in the core-level
spectra of transition metals. The effect, however, was rather small (few {\%})
because of the limited resolution of the experiment. It was later shown that
dichroic effects are also obtained using linearly or even unpolarized
photons~\cite{Ven93,Getz94}. The observed intensity differences in photoemission
are essentially a phenomenon specific to angular-resolved measurements, and
therefore, these have been termed as magnetic circular dichroism in the angular
distribution (MCDAD)~\cite{Cher95,VdL95,Hill96}. Recently, Kozina {\it et
al}~\cite{Kozina:2011} reported strong dichroism in the HAXPES of ferromagnetic
CoFe and Co$_2$FeAl films in exchange bias structure using MnIr, these being
typical materials used in tunnel magnetoresistive devices. Noteworthy, even the
Co 3p peak, which is overlapped by the Ir 4f peak, could be identified by this
technique thanks to the magnetic dichroism effect in that energy
position.~\cite{Kozina:2011} These results have proved the reliability and power
of HAXPES in the study of magnetic properties in buried layers.

This report presents a study of the angular and energy dependence of the magnetic
dichroism in photoemission using hard X-ray excitation. The dichroism of
exchange-biased structures with epitaxially grown ferromagnetic layers of CoFe and
Co$_2$FeAl is investigated. For this purpose the recently installed wide-angle lens
system of the HAXPES end-station of beamline BL47XU of Spring-8 (Japan) was used. The
lens system has the ability to correct spherical aberrations over wide aperture
angles, strongly enhancing the acceptance angle of the high-energy VG Scienta
R4000-HV hemispherical analyzer from about $12^{\circ}$ to about $60^{\circ}$.
\cite{Matsuda:2004,Matsuda:2005}

\section{Theory of magnetic circular dichroism in the angular distribution
of photoelectrons: MCDAD}

Theoretical atomic single-particle models were quite successful in describing,
explaining, and predicting many aspects of magnetic dichroism. Cherepkov {\it et
al}~\cite{Cher95} elaborated the general formalism for the dichroism in
photoemission excited by circularly, linearly, and unpolarized radiation. They
showed that MCDAD is very sensitive to the geometry of the experiment and
depends strongly on the relative orientation between the magnetization,
helicity, and momentum of the excited electrons. The maximum effect is obtained
when the magnetization and helicity vectors are parallel; the effect decreases
with an increase in the angle between these vectors. The electronic states in
solids usually do not carry a spherical or axial symmetry as in free atoms but
have to follow the symmetry of the crystal~\cite{FKC02}. The angular
distribution $I^j({\mathbf{k}},{\bf{n}})$ of the photoemitted electrons -- as
derived for example in Reference~\cite{Cher95} for the case of axially symmetric
polarized atoms -- has to  account for the non- diagonal density matrix
$\rho_{NM_N'}^{\bf{n}}$~\cite{BLU81}. This leads to the following equation for
the case of a non-axial symmetry:

\begin{equation*}
  \begin{array}{ll}
	I^j ( {\mathbf{k}},{\bf{n}} )  = \dfrac{c_\sigma }{[l]}\sqrt{\dfrac{3\left[ j\right] }{4\pi}}
                              \sum\limits_{\kappa ,L}
                              \sum\limits_N\left[ N \right]^{1/2} C_{\kappa LN}^j  \\
                              \times
                              \sum\limits_{x,M}
                              \sum\limits_{M_N,M_N'}
                              \rho_{\kappa x}^\gamma
                              \rho_{NM_N'}^{\bf{n}}(j) Y_{LM}^{*}(\mathbf{k})
                              D_{M_NM'_N}^N(\Omega)
                              \left(
                              \begin{array}{ccc}
	                               \kappa & L & N \\
	                               x      & M & M_N
                              \end{array}
                              \right)
  \end{array}
  \label{eq:pe}
\end{equation*}

\begin{figure}[H]
  \includegraphics[width=6cm]{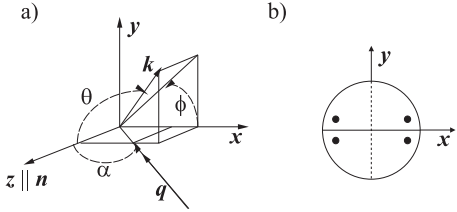}
  \caption{(a)~The coordinate system used for the investigation of photoemission.
           ${\bf k}(\theta,\phi)$ is the electron momentum, $q$ is the photon beam and $n$ is the principal axis of alignment.
           $\theta$ and $\phi$ are the angles defining the direction of the outgoing photoelectrons.
           $\alpha$ is the angle of photon incidence (in the XZ plane) as defined in optics.
           It is seen that the angle describing the photon propagation in spherical coordinates is given by $\Theta_q=\alpha+\pi$.
           The direction of the $z$ axis corresponds to the quantization axis ${\bf n}$.
           (b)~The direction of the in-plane axes $x$ and $y$ is illustrated for an object with $C_{2v}$ symmetry.
           }
\label{fig:vect}
\end{figure}

$l$ and $j$ are the orbital and the total angular momentum of an electron in the
initial state. $C_{\kappa LN}^j$ are the dynamic parameters derived from the
radial matrix elements and $\rho _{\kappa x}^\gamma $ are photon state
multipoles~\cite{BLU81}. $D_{mm_j}^j(\Omega)$ is the Wigner rotation matrix with
$\Omega$ being the set of Euler angles describing the rotation from the
laboratory to the atomic coordinate frame. The direction of the electron
momentum $\stackrel{\rightarrow}{k} ={\bf k}(\theta,\phi)$ is defined by the
angles $\theta$ and $\phi$ (see Figure~\ref{fig:vect}). Finally, $c_\sigma$ is
a photon-energy ($h\nu$) dependent constant: $c_\sigma =\frac{4\pi
^2\alpha\ h\nu}{3}$ where $\alpha$ is the fine structure constant.

This formalism can also be used to consider open shell atoms and the multiplets
resulting from the interaction between the core states and the open shell
valence states. In that case, the dynamic parameters $C_{J\kappa LN}^j$ have to
be calculated for the appropriate coupling scheme ($jj$, LSJ, or intermediate)
with the single particle quantum numbers $j,m$ being replaced by those ($J,M$)
describing the complete atomic state~\cite{Cher95}. In that case, the dynamic
parameter will redistribute the single-electron results in a particular way over
the states of a multiplet (see
References~\cite{TL91a,TholVdL91,TL94a,TholVdL94,Fec01}).

The state multipoles of the $s$ and $p$ states that define the intensity and the
sign and magnitude of the dichroism are summarized in
Table~\ref{tab:statemultipoles01}. Note that the state multipoles are
independent of the orbital angular momentum $L$, they depend only on the total
angular momentum $J$ and its projection $M_J$.

\begin{table}[H]
\centering
\caption{State multipoles of $\left|L, J\right> = \left|0, 1/2\right>$,
         $\left|1, 1/2\right>$, $\left|1, 3/2\right>$, and $\left|2, 3/2\right>$ states.}
    \begin{tabular}{l|cc|cccc}
        $J$          & $\frac{1}{2}$        &                       & $\frac{3}{2}$ & & & \\
        $M_J$        & $+\frac{1}{2}$       & $-\frac{1}{2}$        & $+\frac{3}{2}$ & $+\frac{1}{2}$ & $-\frac{1}{2}$ & $-\frac{3}{2}$\\
        \noalign{\smallskip}\hline\noalign{\smallskip}
        $\rho_{00}$  & $\frac{1}{\sqrt{2}}$ & $\frac{1}{\sqrt{2}}$  & $\frac{1}{2}$ & $\frac{1}{2}$  & $\frac{1}{2}$  & $\frac{1}{2}$ \\
        $\rho_{10}$  & $\frac{1}{\sqrt{2}}$ & $-\frac{1}{\sqrt{2}}$ & $\frac{3}{2\sqrt{5}}$ & $\frac{1}{2\sqrt{5}}$ & $-\frac{1}{2\sqrt{5}}$ &  $-\frac{3}{2\sqrt{5}}$ \\
        $\rho_{20}$  & -                    & -                     & $\frac{1}{2}$ & $-\frac{1}{2}$ & $-\frac{1}{2}$ & $\frac{1}{2}$ \\
        $\rho_{30}$  & -                    & -                     & $\frac{1}{2\sqrt{5}}$ & $-\frac{3}{2\sqrt{5}}$ & $\frac{3}{2\sqrt{5}}$ &  $-\frac{1}{2\sqrt{5}}$ \\
    \end{tabular}
    \label{tab:statemultipoles01}
\end{table}

\subsection{Equations for grazing incidence geometry.}
\label{sec:eqexp}

In the following, let us consider the special case of geometry of the
experiments as described below. A small deviation caused by the finite angle
($1^\circ$) of the X-rays with respect to the surface will be neglected. The
photons are impinging in the $x-z$ plane with unit vector of the photon momentum
$\hat{{q}} = (-\cos(\alpha),-\sin(\alpha),0)$. At such a grazing incidence with
$\alpha=\pi/2$ it becomes $\hat{{q}} = (-1,0,0)$. The electrons are observed in
the direction perpendicular to the photon beam ($\theta=\frac{\pi}{2}-\alpha$)
with the momentum $\hat{{k}} = (-\sin(\theta),0,\cos(\theta))$. At a photon
incidence of $\alpha=\pi/2$ and normal emission it becomes $\hat{{k}} = (0,0,1)$. (Compare also
Figures~\ref{fig:vect} and~\ref{fig:samples}.)

Now examine the case: $\stackrel{\rightarrow}{n} \rightarrow -
\stackrel{\rightarrow}{n}$ where the magnetic dichroism emerges from a switching
of the direction of magnetization with the initial direction
$\stackrel{\rightarrow}{n}=(1,0,0)$ that is along the $x$-axis. Applying
Equation~(\ref{eq:pe}) and the state multipoles of
Table~\ref{tab:statemultipoles01} the circular magnetic dichroism in the
experimental geometry is given for $p$-states ($J=1/2, 3/2$) by the simple
equation:

\begin{equation}
  \begin{array}{ll}
   MCDAD^{\sigma\pm}(p_J) = & \\
    \mp \rho_{10} \left( \sqrt{\frac{1}{3}} C_{JkLN}^{(1, 0, 1)}
                                                - \sqrt{\frac{1}{15}} C_{JkLN}^{(1, 2, 1)}
                                                + \sqrt{\frac{3}{10}} C_{JkLN}^{(1, 2, 1)} \cos^2(\theta) \right)
  \end{array}
\label{eq:cmdad_p}
\end{equation}

The magnetic circular dichroism in the angular distribution (MCDAD) for opposite
helicity of the photons has an opposite sign. The equations for the $p_{1/2}$
and $p_{3/2}$ states are the same. The magnitude differs, however, because of
the differences in the state multipoles $\rho_{10}$ and dynamical parameters
$C_{JkLN}$. The $\cos^2$ function in Equation~(\ref{eq:cmdad_p}) leads to a
rather smooth variation of the magnetic dichroism in the geometry used for the
present experiment.

\section{Experimental details} 

\begin{figure}[H]
\centering
\includegraphics[width=8cm, clip]{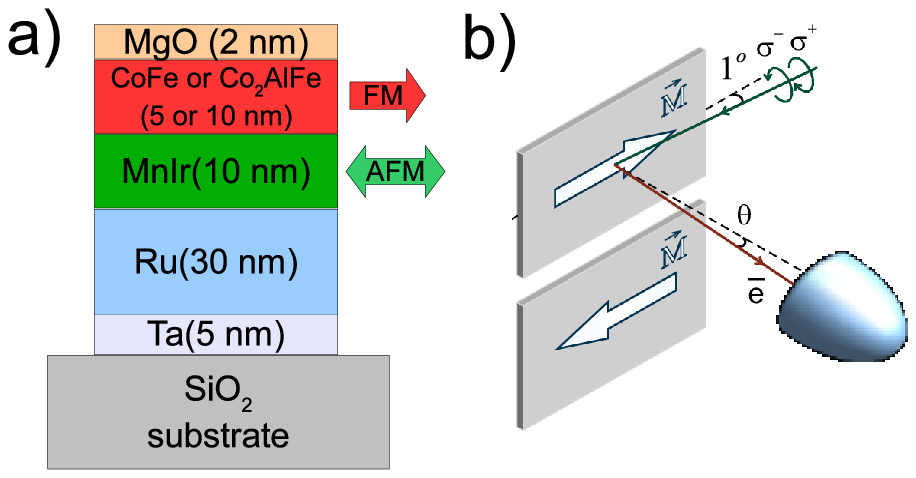}
\caption{(Color online) (a) Sketch of the exchange-biased films used in the dichroism experiments. \newline
          (b) Scheme of the experimental geometry.
          The incidence angle of the circularly polarized photons was fixed to $1^{\circ}$.
          X-rays of opposite helicity ($\sigma^+$ and $\sigma^-$) were provided by a
          phase retarder. Identical samples with opposite directions of magnetization are used
          to avoid the necessity of magnetization reversal by external high magnetic fields.
          The in-plane magnetization $M$ is nearly parallel to the beam axis in order to maximize
          the dichroism. The photoelectrons are detected over a large range of emission angles $\theta$. }
\label{fig:samples}
\end{figure}

DC/RF magnetron sputtering was used for the preparation of half of the stack for
magnetic tunnel junctions. All films were deposited at room temperature. The
argon process pressure was set to $1.5\times 10^{-3}$~mbar in the sputtering
system where the base pressure was about 10$^{-7}$~mbar. The layers were
deposited on thermally oxidized SiO$_2$ substrates in the following order:\\
i) Ta(5) / Ru(30) / MnIr(10) / CoFe(10) / MgO(2)  \\
ii) Ta(5) / Ru(30) / MnIr(10) / Co$_2$FeAl(10) / MgO(2) \\
The numbers in brackets give the corresponding layer thickness in nm. The stack
sequence is sketched in Figure~\ref{fig:samples}a. The samples were annealed at
$275^{\circ}$C for 10~min in vacuum ($5\times 10^{-2}$~Pa) in a magnetic field
of 0.1~T to provide exchange biasing of the ferromagnetic layer through the
MnIr/CoFe(Co$_2$FeAl) interface (see also~\cite{Mar07}), in such a way the CoFe
or Co$_2$FeAl layers are kept magnetized in preset directions.

The HAXPES experiments with an excitation energy of 7.940~keV were performed at
beamline BL47-XU of SPring-8~\cite{Kob09}. The energy and angular distribution
of the photoexcited electrons was analyzed using a high energy VG Scienta R4000-HV
hemispherical analyzer, in front of which a wide-acceptance objective
lens~\cite{Matsuda:2004,Matsuda:2005} was placed. The effective acceptance angle
was enlarged to about $60^{\circ}$ with an angular resolution of $1^{\circ}$.
The overall energy resolution was about 250~meV depending on pass
energy and setting of the monochromator. The angle between the electron
spectrometer and the photon propagation was fixed at $90^{\circ}$. The impinging
angle was set to $1^{\circ}$ (resulting in $\theta=-89^{\circ}$) in order to
ensure that the polarization vector of the circularly polarized photons is
nearly parallel ($\sigma^-$) or antiparallel ($\sigma^+$) to the in- plane
magnetization $M^+$ (see Fig~1b). The sign of the magnetization was varied by
mounting identical samples with opposite directions of magnetization ($M^+$,
$M^-$). This allowed to probe the dichroism by varying both the direction of
magnetization and the direction of helicity. The vertical spot size on the
sample is  30~$\mu$m, while in horizontal direction, along the entrance slit of
the analyzer, the spot was stretched to approximately 7~mm. The polarization of
the incident photons was varied using an in-vacuum phase retarder based on a
600-$\mu$m-thick diamond crystal with (220) orientation~\cite{SKM98}. The direct
beam is linearly polarized with $P_p=0.99$. Using the phase retarder, the degree
of circular polarization is set such that $P_{c} > 0.9$. All measurement were
performed at room temperature (\~300~K). The circular dichroism is usually
characterized by an asymmetry that is defined as the ratio of the difference
between the intensities $I^+$ and $I^-$ and their sum, $A=(I^+- I^-)/(I^++I^-)$,
where $I^+$ corresponds to $\sigma^+$ ($e_x+i\cdot e_y$) and $I^-$ to $\sigma^-$
($e_x-i\cdot e_y$) type helicity. Here, $e_{x,y}$ are the $x$ and $y$ components
of the complex electric field vector describing circularly polarized photons
propagating along the $z$ axis. It is difficult, however, to establish a
standard procedure for background subtraction. Moreover, small noise in the
numerator of $A$ can produce strongly divergent values when the denominator goes
to zero after background subtraction. For this reason, the effects of the
magnetic dichroism are simply shown as difference spectra $\Delta I= I^+- I^-$.

\section{Results and discussion} 

In the following, the magnetic circular dichroism in the photoelectron spectra
of the $2p$ core level of Co and Fe will be presented and discussed. The spectra
were taken from exchange biased thin CoFe and Co$_2$FeAl films, two materials
that are widely used in spintronic devices. Firstly, results from the angular
integrated mode will be reported to explain the basic effect. In the second
part, details of the angular resolved measurements are presented. Finally,
angular integrated magnetic dichroism in emission from the valence band will be presented.

\subsection{Angular integrated magnetic dichroism.}

\begin{figure}[H]
  \centering
  \includegraphics[width=8cm]{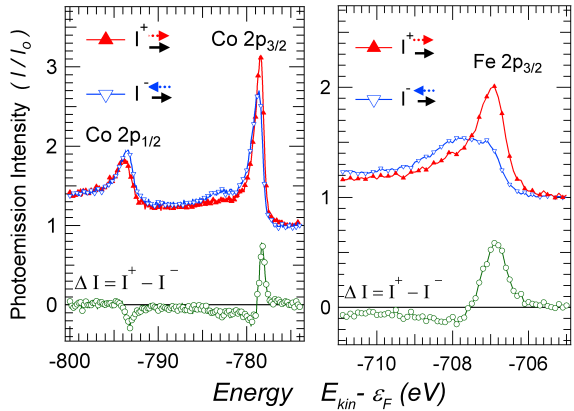}
  \caption{(Color online) Polarization-dependent photoelectron spectra of
            the Co $2p$ (left panel) and Fe $2p_{3/2}$ (right panel)
            core level emission from 10~nm-thick CoFe film on top of the MnIr exchange-biasing layer.
            $I_0$ represents the background intensity anteceding the lower energy tail.
            Dotted and solid arrows indicated the x-ray helicity and sample magnetization respectively.
            The difference spectra are represented by the open circles. }
\label{fig_CoFe}
\end{figure}

As starting point, the regular -- that is angular integrated, polarization
dependent -- photoemission spectra of the exchange-biased CoFe film were
measured using the \emph{transmission mode} of the spectrometer. The results are
in well agreement to those reported earlier for similar
systems~\cite{Kozina:2011}. As shown in Figure~\ref{fig_CoFe}, the $2p$ core
level spectra of Co (left panel) and Fe (right panel) exhibit a pronounced
difference when taken with photons of opposite helicity for a fixed direction of
magnetization. The pure difference $\Delta I=I^+ - I^-$ shown in
Figure~\ref{fig_CoFe} contains all characteristic features of the magnetic
circular dichroism.

As one can see from Figure~\ref{fig_CoFe}a, the spin-orbit splitting of the Co
$2p$ states is clearly resolved. The dichroism changes its sign across the Co
$2p$ spectrum in the sequence: $+ - - +$ when going from $p_{3/2}$ to $p_{1/2}$.
This appears to be characteristic of a Zeemann-type $m_j$ sub-level ordering as
a consequence of the signs of the state multipoles $\rho_{10}$ given in
Table~\ref{tab:statemultipoles01}. The details of the MCDAD reveal, however,
that the situation is more complicated. There are clear hints on multiplett
splittings seen in the spectra. Close below the Co $p_{3/2}$ excitation, a low
intense satellite is seen that exhibits clearly a dichroic signal. The asymmetry
of this satellite is clearly higher (note the low intensity above background)
compared to that at the maximum of the Co $p_{3/2}$ intensity. Also, the
dichroism in the Fe $2p$ spectrum does not vanish in the region between the
spinorbit doublet, what complicates the analysis in terms of an asymmetry. It is
a clear indicator for multiplet states that spread over the hole Fe $2p$
spectrum. The splitting of the Fe $2p_{3/2}$ is clearly more pronounced compared
to Co $2p_{3/2}$. The details of the splitting an the dichroism have been discussed more
detailed already in Reference~\cite{Kozina:2011}.

\subsection{Angular resolved magnetic dichroism.}

In the next step, the angular distribution of the magnetic circular dichroism
was investigated. In the previous experiments~\cite{Kozina:2011}, the angular
resolution was about $\pm5^\circ$ and the electron detection was performed for
normal emission only. Here, details of the magnetic dichroism will be studied
over a $60^\circ$ range of angles with a resolution of about $0.2^{\circ}$ to
$2^{\circ}$.

\begin{figure}[H]
  \centering
  \includegraphics[width=7.6cm]{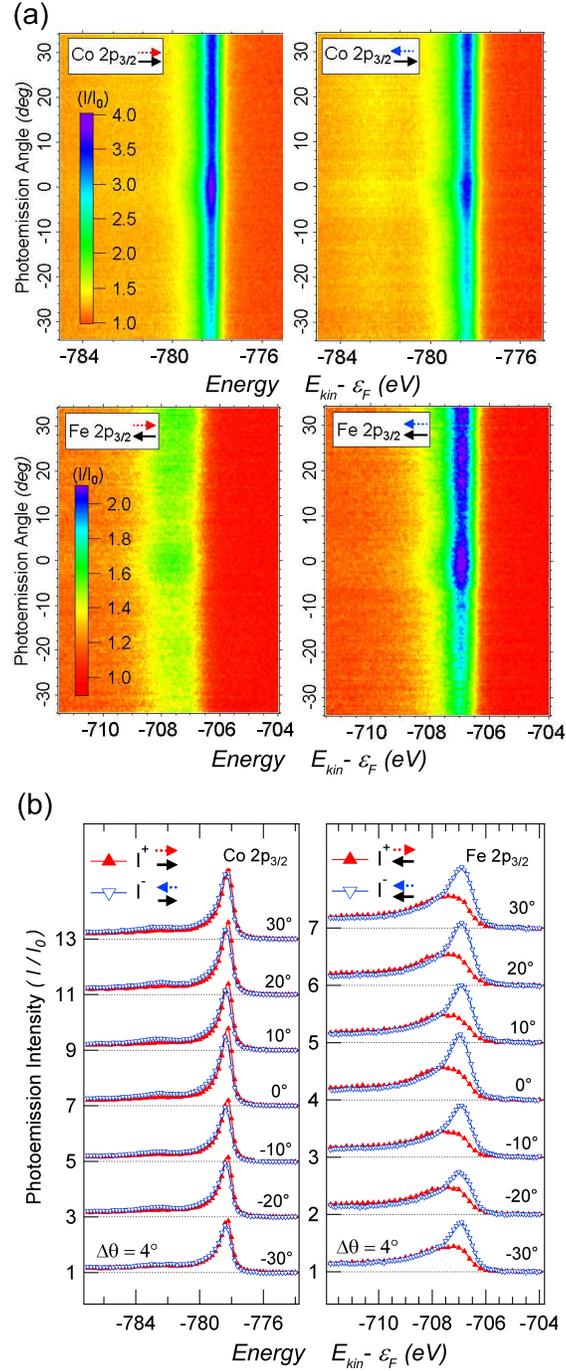}
  \caption{(Color online) Angular distribution of polarization-dependent photoelectron spectra
            of the Co and Fe $2p_{3/2}$ core level emission from 10~nm thick Co$_{2}$FeAl film
            on top of an MnIr exchange-biasing layer. $I_0$ represents the background intensity
            anteceding the lower energy tail. Dotted and solid arrows indicated the x-ray helicity
            and sample magnetization respectively. The bottom panel shows the EDC for both peaks and polarizations.
            Curves of different photoemission angles are integrated with a slice of 4$^\circ$ and are
            vertically offset from the zero value for clarity.}
\label{fig_AR}
\end{figure}

Figure~\ref{fig_AR} shows the $2p$ core level spectra of Co and Fe taken from
the exchange biased Co$_2$FeAl film. The spectra were measured in the
\emph{angular mode}, which takes advantage of the wide-angle condenser lens
system allowing the electrons to be collected over emission angles from
$-33^{\circ}$ to $+33^{\circ}$ around normal emission. At the spectrometer, each
channel of the detector covers an angle slice of $0.167^{\circ}$ width. In general,
measurements of the angular distribution of the photoelectrons using hard X-ray
source and phase-retarder are very time consuming. The electron flux impinging
each angular channel of the detector is very small resulting in a low
accumulation rate and poor statistics compared to angular integrated spectra
taken with the same integration time. For practical purposes, the spectra must
be integrated over several channels to be analyzed.

The energy distribution curves (EDC) are shown in the bottom panel of
Figure~\ref{fig_AR}. The spectra are summed up over slices of $\pm2^{\circ}$ (12
channels) about the photoemission direction indicated by the label. No striking
differences are noticed in spectral shape collected at different emission
angles, but a strong dependence on the photon polarization is noticed in both,
Co and Fe, $2p_{3/2}$ peaks. In comparison with the CoFe spectra (see
Figure~\ref{fig_CoFe}a), the Co $2p_{3/2}$ has a shift to lower binding energies
and exhibits a smaller dichroism.  The Fe $2p$ peaks in Co$_2$FeAl are similar
to those measured from CoFe (see Figure~\ref{fig_CoFe}b), where the spectral
shapes for $I^+$ and $I^-$ are exchanged due to the reversal of magnetization
between the two samples.

\begin{figure}[H]
  \centering
  \includegraphics[width=7.6cm]{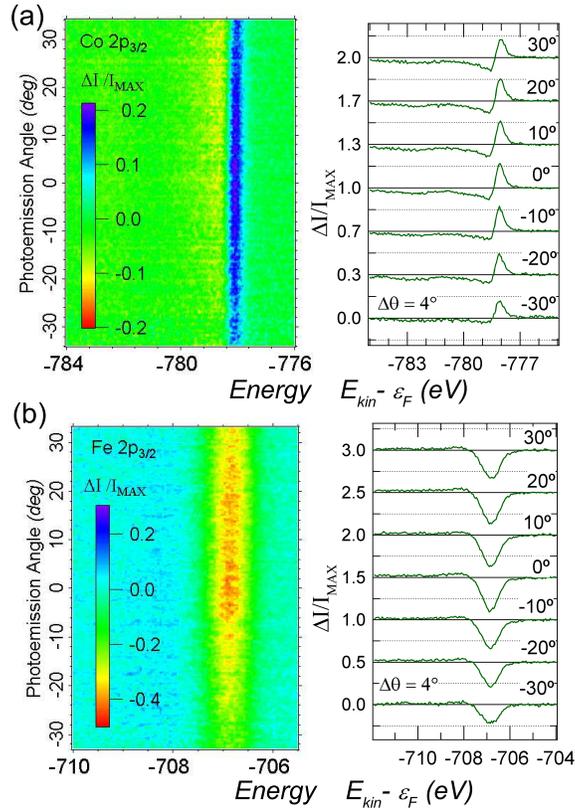}
  \caption{(Color online) Angular distribution of circular dichroism in the Co $2p_{3/2}$ (a)
            and Fe  $2p_{3/2}$ (b) core-level photoemission from 10nm-thick Co$_{2}$FeAl film.
            $\Delta I$ is shown normalized by $I_{MAX}(\theta)$ which is the peak dispersion
            of the average photoemission (see text).
            Horizontal line-profiles are shown on right panels; curves for different photoemission
            angles are vertically offset from the zero value for clarity.}
\label{fig_diff}
\end{figure}

\begin{figure}[H]
  \centering
  \includegraphics[width=7cm]{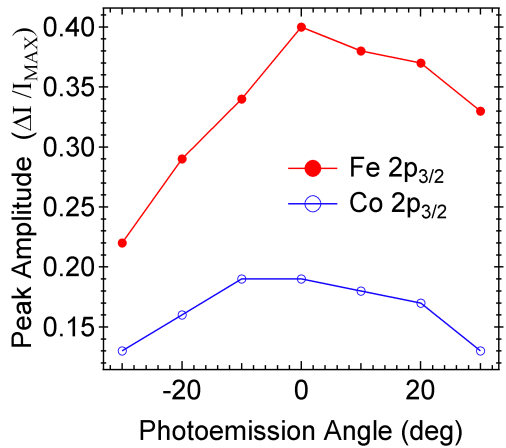}
  \caption{(Color online) Peak  dispersion of $\Delta I/I_{MAX}(\theta)$ for the Co and Fe $2p_{3/2}$
           states of the Co$_{2}$FeAl film.}
\label{fig_peakdisp}
\end{figure}

The angular dependence of the circular dichroism is shown in all details in
Figure~\ref{fig_diff}. The dichroism $\Delta I$ is normalized by
$I_{MAX}(\theta)$, which represents the peak dispersion of the average spectra
$I^{avg}=(I^+ - I^-)/2$. In this way, the angular dependence of the circular
dichroism is free of structural contributions (e.g.: due to the forward
scattering effect\cite{Poon:1984,Egelhoff:1990}) and it results uniquely from
the magnetic properties and symmetry of the electronic states. Note that, for
both Fe and Co $2p_{3/2}$ peaks, there is an enhancement of dichroism about the
normal direction. This is clearer seen from Figure~\ref{fig_peakdisp}. The $\cos^2$
function in Equation~(\ref{eq:cmdad_p}) leads to a maximum close to normal
emission as is observed in the experiment. The small variation of the dichroism
with angle of emission reveals that an atomic like description of the MCDAD as
given above is sufficient to explain its most important features of the angular
distribution. The complete energy dependence is, indeed, more complicated and is
somehow hidden in the dynamical parameters $C^j_{JkLN}$ of
Equations~(\ref{eq:cmdad_p}) and~(\ref{eq:pe}). As mentioned above, these
parameters include all possible multiplett and crystal field effects.

\subsection{Magnetic dichroism in valence band spectroscopy.}

As already demonstrated in Reference~\cite{Kozina:2011}, the magnetic dichroism
is not only present in the photoelectron spectra of core level but also in
valence band spectra. Different from excitation with UV photons~\cite{SHS91,KSc01} and
low kinetic energies, HAXPES does not
reveal emission from single bands but averages in $k$-space and is thus
proportional to the density of states, indeed convoluted with the cross sections
of the contributing states. It is not clear from beginning whether or not this
allows to detect any magnetic circular dichroism, in particular when using an
angular integrating mode.

\begin{figure}[H]
  \centering
  \includegraphics[width=7cm]{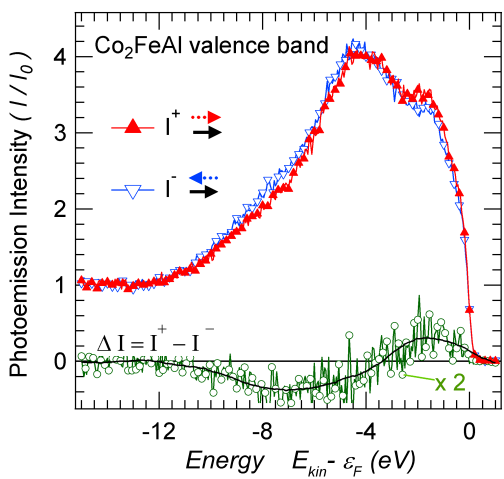}
  \caption{(Color online) Polarization-dependent photoelectron valence band spectra
            from 10nm-thick Co$_{2}$FeAl film on top of the MnIr exchange-biasing layer.
            Dotted and solid arrows indicated the X-ray helicity and sample magnetization respectively.
            The difference of the spectra taken with the opposite helicity of light is
            represented by open-circles and is shown enlarged by a factor of two.}
\label{fig_vbCFA}
\end{figure}

Figure~\ref{fig_vbCFA} shows the polarization dependence of the valence band
spectra from the Co$_2$FeAl multilayer. It should be stressed that the valence
band spectra have important contributions from the MnIr layer as well as from
other underlying layers, nevertheless, a strong effect is still visible in the
spectra near the Fermi energy when changing the helicity of the X-rays.
Tentatively, this difference in the spectra is a consequence of a depletion of
the occupied minority spin states close to the Fermi edge of the Co$_2$FeAl
layer.

Here the spectra are integrated in $k$-space over the initial states in the
Brillouin zone as well as over a wide range ($\pm33^\circ$) of emission angles.
From the spectra and dichroism shown in Figure~\ref{fig_vbCFA} it is obvious
that magnetic circular dichroism is also observed in such integrated spectra.
This demonstrates that magnetic dichroism can be observed even in the bulk
sensitive spectroscopy of the total density of states at high kinetic energies
($\approx8$~keV).

\section{Summary and Conclusion} 

In summary, the feasibility to measure simultaneously the angular and energy
distribution of the magnetic circular dichroism in photoelectron spectroscopy
was demonstrated by use of a wide-angle lens setup installed at the HAXPES end-station
of the beamline BL47XU at Spring-8. Bulk-sensitive HAXPES-MCDAD was used to
image the electronic states and to study the magnetic response from buried
layers. CoFe and Co$_2$FeAl ferromagnetic films grown on MnIr exchange-biased
magnetic layers buried beneath MgO layers were analyzed. A strong dichroism was
observed not only for core level but also in the valence band spectra of
Co$_2$FeAl.

Recently, first spin resolved HAXPES experiments were
reported~\cite{GSF12,SKF12}. The much higher intensities (by three orders of
magnitude) in spin integrated spectra makes HAXPES-MCDAD attractive for the
studies of magnetic bulk materials and buried thin films in material science.
HAXPES-MCDAD is particularly still the method of choice in valence band spectroscopy
with low cross sections at high kinetic energies where the intensities are still
too low for spin-resolved methods also with large angles of acceptance at the spectrometer.

Overall, the high bulk sensitivity of angular resolved HAXPES combined with
circularly polarized photons will have a major impact on the study of the
magnetic phenomena of deeply buried magnetic materials. It will allow an
element-specific study of the magnetism of buried layers and make feasible the
investigation of the properties of magnetic layers not only at the surface but
also at buried interfaces.

\section*{Acknowledgments}

The authors thank E. Ikenaga (Spring-8) for help with the experiments.
D.~E. thanks the group of G. Reiss (University of Bielefeld) for help
with the sample production. Financial
support by {\it Deutsche Forschungsgemeinschaft DFG} and the Strategic
International Cooperative Program of the {\it Japan Science and Technology
Agency JST} (DFG-JST) (Project P~1.3-A of research unit FOR 1464: {\it
ASPIMATT}) is gratefully acknowledged. The HAXPES experiment was performed at
BL47XU of SPring-8 with approval of JASRI (Proposal No.~2012A0043).

\section*{References}
\bibliographystyle{elsarticle-num}

\begin{thebibliography}{10}
\expandafter\ifx\csname url\endcsname\relax
  \def\url#1{\texttt{#1}}\fi
\expandafter\ifx\csname urlprefix\endcsname\relax\def\urlprefix{URL }\fi
\expandafter\ifx\csname href\endcsname\relax
  \def\href#1#2{#2} \def\path#1{#1}\fi

\bibitem{Lind74}
I.~Lindau, P.~Pianetta, S.~Doniach, W.~E. Spicer, Nature 250.

\bibitem{Koba03}
K.~Kobayashi, M.~Yabashi, Y.~Takata, T.~Tokushima, S.~Shin, K.~Tamasaku,
  D.~Miwa, T.~Ishikawa, H.~Nohira, T.~Hattori, Y.~Sugita, O.~N.~A. Sakai,
  S.~Zaima, Appl. Phys. Lett. 83 (2003) 1005.

\bibitem{Suga09}
S.~Suga, A.~Sekiyama, Eur. Phys. J. Special Topics 169 (2009) 227.

\bibitem{FBG08}
G.~H. Fecher, B.~Balke, A.~Gloskowskii, S.~Ouardi, C.~Felser, T.~Ishikawa,
  M.~Yamamoto, Y.~Yamashita, H.~Yoshikawa, S.~Ueda, K.~Kobayashi, Appl. Phys.
  Lett. 92 (2008) 193513.

\bibitem{Koz10}
X.~Kozina, S.~Ouardi, B.~Balke, G.~Stryganyuk, G.~H. Fecher, C.~Felser,
  S.~Ikeda, H.~Ohno, E.~Ikenaga, Appl. Phys. Lett. 96 (2010) 072105.

\bibitem{Ueda08}
S.~Ueda, H.~Tanaka, J.~Takaobushi, E.~Ikenaga, J.-J. Kim, M.~Kobata, T.~Kawai,
  H.~Osawa, N.~Kawamura, M.~Suzuki, K.~Kobayashi, Appl. Phys. Exp. 1 (2008)
  077003.

\bibitem{Ouardi11}
S.~Ouardi, G.~H. Fecher, X.~Kozina, G.~Stryganyuk, B.~Balke, C.~Felser,
  E.~Ikenaga, T.~Sugiyama, N.~Kawamura, M.~Suzuki, K.~Kobayashi, Phys. Rev.
  Lett. 107 (2011) 036402.

\bibitem{Baum90}
L.~Baumgarten, C.~M. Schneider, H.~Petersen, S.~F, J.~Kirschner, Phys. Rev.
  Lett. 65 (1990) 492.

\bibitem{Ven93}
D.~Venus, Phys. Rev. B 48 (1993) 6144.

\bibitem{Getz94}
M.~Getzlaff, C.~Ostertag, G.~H. Fecher, N.~A. Cherepkov, G.~Sch{\"o}nhense,
  Phys. Rev. Lett. 73 (1994) 3030.

\bibitem{Cher95}
N.~A. Cherepkov, V.~V. Kuznetsov, V.~A. Verbitskii, J. Phys. B: At. Mol. Opt.
  Phys. 28 (1995) 1221--1239.

\bibitem{VdL95}
G.~van~der Laan, B.~T. Thole, Phys. Rev. B 52 (1995) 15355.

\bibitem{Hill96}
F.~U. Hillebrecht, C.~Roth, H.~B. Rose, W.~G. Park, E.~Kisker, N.~A. Cherepkov,
  Phys. Rev. B 53 (1996) 12182.

\bibitem{Kozina:2011}
X.~Kozina, G.~H. Fecher, G.~Stryganyuk, S.~Ouardi, B.~Balke, C.~Felser,
  G.~Sch\"onhense, E.~Ikenaga, T.~Sugiyama, N.~Kawamura, M.~Suzuki, T.~Taira,
  T.~Uemura, M.~Yamamoto, H.~Sukegawa, W.~Wang, K.~Inomata, K.~Kobayashi, Phys.
  Rev. B 84 (2011) 054449.

\bibitem{Matsuda:2004}
H.~Matsuda, H.~Daimon, Japan Patent: PCT/jp2004/016602 Japan 208926 (2004).

\bibitem{Matsuda:2005}
H.~Matsuda, H.~Daimon, M.~Kato, M.~Kudo, Phys. Rev. E 71 (2005) 066503.

\bibitem{FKC02}
G.~H. Fecher, V.~V. Kuznetsov, N.~A. Cherepkov, G.~Sch\"onhense, J. Electron
  Spectrosc. Relat. Phenom. 122 (2002) 157.

\bibitem{BLU81}
K.~Blum, Density Matrix Theory and Application, Plenum, New York, 1981.

\bibitem{TL91a}
B.~T. Thole, G.~v.~d. Laan, Phys. Rev. Lett. 67 (1991) 3306.

\bibitem{TholVdL91}
B.~T. Thole, G.~van~der Laan, Phys. Rev. B 44 (1991) 12424.

\bibitem{TL94a}
B.~T. Thole, G.~v.~d. Laan, Phys. Rev. B 50 (1994) 11474.

\bibitem{TholVdL94}
B.~T. Thole, G.~van~der Laan, Phys. Rev. B 49 (1994) 9613.

\bibitem{Fec01}
G.~H. Fecher, J. Elect. Spectros. Rel. Phenom. 114-116 (2001) 1165.

\bibitem{Mar07}
T.~Marukame, T.~Ishikawa, S.~Hakamata, K.-i. Matsuda, T.~Uemura, , M.~Yamamoto,
  Appl. Phys. Lett. 90 (2007) 012508.

\bibitem{Kob09}
K.~Kobayashi, Nucl. Instr. Meth. Phys. Res. A 601 (2009) 32 -- 47.

\bibitem{SKM98}
M.~Suzuki, N.~Kawamura, M.~Mizukami, A.~Urata, H.~Maruyama, S.~Goto,
  T.~Ishikawa, Jpn. J. Appl. Phys. 37 (1998) L1488.

\bibitem{Poon:1984}
H.~C. Poon, S.~Y. Tong, Phys. Rev. B 30 (1984) 6211.

\bibitem{Egelhoff:1990}
J.~W.~F.~Egelhoff, I.~Jacob, J.~M. Rudd, J.~F. Cochran, B.~Heinrich, J. Vac.
  Sci. Technol. A 8~(3) (1990) 1582.

\bibitem{SHS91}
C.~M. Schneider, M.~S. Hammond, P.~Schuster, A.~Cebollada, R.~Miranda,
  J.~Kirschner, Phys. Rev. B 44 (1991) 12066.

\bibitem{KSc01}
W.~Kuch, C.~M. Schneider, Rep. Prog. Phys. 64 (2001) 147.

\bibitem{GSF12}
A.~Gloskovskii, G.~Stryganyuk, G.~H. Fecher, C.~Felser, S.~Thiess,
  H.~Schulz-Ritter, W.~Drube, G.~Berner, M.~Sing, R.~Claessen, M.~Yamamoto, J.
  Electron Spectrosc. Relat. Phenom. 185 (2012) 47.

\bibitem{SKF12}
G.~Stryganyuk, X.~Kozina, G.~H. Fecher, S.~Ouardi, S.~Chadov, C.~Felser,
  G.~Sch{\"o}nhense, P.~Lushchyk, A.~Oelsner, P.~Bernhard, E.~Ikenaga,
  T.~Sugiyama, H.~Sukegawa, Z.~Wen, K.~Inomata, K.~Kobayashi, Jpn. J. Appl.
  Phys. 51 (2012) 016602.

\end{thebibliography}

\end{document}